\documentclass[runningheads]{llncs}

\usepackage[T1]{fontenc}
\usepackage{graphicx}
\usepackage{enumerate}
\usepackage{multicol}
\usepackage[frozencache,cachedir=.]{minted} 
\usepackage{hyperref}
\hypersetup{bookmarks,bookmarksopen,bookmarksdepth=2}
\usepackage{bookmark}

\begin{document}
\pagenumbering{arabic} 
\title{Data Privacy Vocabulary (DPV) —  Version 2.0}
\author{
    Harshvardhan J. Pandit\inst{1} \and
    Beatriz Esteves\inst{2}  \and
    Georg P. Krog\inst{3} \and
    Paul Ryan\inst{4} \and
    Delaram Golpayegani\inst{5} \and
    Julian Flake\inst{6}
}
\authorrunning{H. J. Pandit et al.}
%
\institute{
    ADAPT Centre, Dublin City University, Dublin, Ireland, \email{me@harshp.com} \and
    IDLab, Ghent University —  imec, 
    Ghent, Belgium, \email{beatriz.esteves@ugent.be} \and
    Signatu AS, Oslo, Norway \and
    ADAPT Centre, Dublin City University, and Uniphar PLC, Dublin, Ireland \and
    ADAPT Centre, Trinity College Dublin, Dublin, Ireland \and
    University of Koblenz, Koblenz, Germany, \email{flake@uni-koblenz.de}
}

\maketitle

\begin{abstract}
The Data Privacy Vocabulary (DPV), developed by the W3C Data Privacy Vocabularies and Controls Community Group (DPVCG), enables the creation of machine-readable, interoperable, and standards-based representations for describing the processing of personal data. The group has also published extensions to the DPV to describe specific applications to support legislative requirements such as the EU's GDPR. The DPV fills a crucial niche in the state of the art by providing a vocabulary that can be embedded and used alongside other existing standards such as W3C ODRL, and which can be customised and extended for adapting to specifics of use-cases or domains. This article describes the version 2 iteration of the DPV in terms of its contents, methodology, current adoptions and uses, and future potential. It also describes the relevance and role of DPV in acting as a common vocabulary to support various regulatory (e.g., EU's DGA and AI Act) and community initiatives (e.g., Solid) emerging across the globe.
\end{abstract}

\section{Introduction}\label{sec:intro}

The modern technological landscape consists of ubiquitous digital devices and services which generate vast amounts of data, which includes sensitive information that raises privacy concerns, as well as requires the protection of data from misuse and cybersecurity threats. Regulations across the globe have been developed or updated to meet this challenge, most notably the European Union's (EU) General Data Protection Regulation (GDPR) \cite{GDPR} in 2016, which requires specific activities to be carried out based on defined norms and requirements, and require documenting governance processes for compliance. 

`Regulatory Technology' (RegTech) has also evolved to provide information management capabilities and automation of tasks to support evolving regulations. However, a key barrier to their effective use is their proprietary nature, non-interoperable information, and lack of standards. As a result, the RegTech landscape for privacy and data protection is fragmented and siloed and lacks any meaningful `metadata' and `semantics' through which information can be reused, e.g., for processes not envisioned by a tool provider.

Based on this background, the SPECIAL H2020 project established the W3C Data Privacy Vocabularies and Controls Community Group\footnote{\url{https://www.w3.org/groups/cg/dpvcg/}} (DPVCG) in 2018, which developed the Data Privacy Vocabulary\footnote{\url{https://w3id.org/dpv}} (DPV) \cite{panetto_creating_2019} as a machine-readable interoperable vocabulary for the exchange of legally relevant `metadata'. Since then, DPV has continued being a state of the art resource that is iteratively updated to match the evolving landscape of regulations and compliance requirements. It has seen high adoption in academic, industrial, and mixed settings, and has been referenced in standards. 

The DPV 1.0 focused on providing a vocabulary for describing personal data activities based on the GDPR. Since then, the world has evolved at a rapid pace with new innovations and regulations --  in particular those related to `increasing sharing of data' such as the EU's Data Governance Act (DGA)~\cite{DGA}, as well as for `regulating AI', such as the EU's Artificial Intelligence Act (AI Act) \cite{AIAct}. To reflect these developments as well as others in domains such as cybersecurity, the DPVCG has updated the DPV and made it capable of representing a wide variety of legally relevant information for `AI and digital regulations'. In this article, we present the design and development of DPV -- version 2.

The rest of the article is structured as follows: Section \ref{sec:req} describes the requirements which guided the development of DPV, Section \ref{sec:related} compares DPV with related work, Section \ref{sec:adoption} describes the known adoptions, Section \ref{sec:overview} provides an overview of DPV, Section \ref{sec:design} describes the methodology and design principles used, and Section \ref{sec:conclusion} concludes the article with a discussion on future work.

\section{Requirements for a Legal Vocabulary}\label{sec:req}

\subsection{Information and Knowledge Modelling}
DPV 1.0 was concerned with modelling the purposes, processing operations, legal bases, personal data categories, technical and organisational measures, and the roles of entities based on the GDPR, and to fill specific gaps~\cite{panetto_creating_2019}:
(1) there were no ontologies to describe and exchange information about activities involving personal data;
(2) there were no agreed upon vocabularies or taxonomies for representing practical applications and uses of personal data, e.g., different purposes such as \textit{Service Provision}, or \textit{Legal Bases} such as \textit{Consent}; and
(3) there were no vocabularies that align the terminology and requirements of regulations for developing and using machine-readable information.

Based on these, DPV 1.0 was developed to provide:
\begin{enumerate}
    \item a formal ontology that defines concepts and relationships;
    \item a (hierarchical) taxonomy that provides \textit{instances} or \textit{specialisations} of the ontological concepts to reflect practical uses and applications; and
    \item a legally relevant modelling of information to support documentation needs for compliance and exchange of information between stakeholders.
\end{enumerate}

For DPV 2.0, the evolving landscape of law and technologies and the initial adoption of DPV led the DPVCG to reconsider the scope of DPV. This resulted in the following additional requirements:
\begin{enumerate}
    \item supporting multiple jurisdictions and laws (e.g., EU — GDPR, US — CCPA);
    \item supporting risk management (e.g., based on ISO 31000 series);
    \item representing use of services and technologies, e.g., cloud services;
    \item describing how AI (as a technology) is used;
    \item providing guidance on use of DPV to meet specific legal requirements; and
    \item providing documentation, examples, and guidance for increasing adoption.
\end{enumerate}

\subsection{Legal and Semantic Extensibility}

Due to the nature of legal frameworks, concepts in DPV must constantly be assessed and possibly modified to support changes in laws and case law. Unlike conventional ontologies, the \textit{meaning} and \textit{semantics} of concepts in DPV relies on legal norms and interpretations which differ across jurisdictions and change with laws and rulings.
The design of DPV therefore requires the possibility for it to be used in a `jurisdiction-independent' manner while also having a mechanism to explicitly assert a specific law, or how its concept applies.
For example, `consent' is a generic concept, whereas `consent according to GDPR' is a jurisdiction and regulation-specific concept. Such \textit{legal customisability} is an important part of DPV's interoperability as it enables different stakeholders to express common requirements (e.g., consent) which can be explicitly asserted or interpreted in relevant contexts (e.g., regulation based on location). 

Semantic interoperability is a cornerstone of DPV given its emphasis on representation and exchange of information between stakeholders. Compared to conventional ontologies, concepts in DPV must be extensible to support practical requirements which do not align neatly with design considerations of only providing classes and properties. For example, a concept `Email' that represents personal data associated with emails cannot be simply modelled as a class as it may be needed as an \textit{instance} to state emails are collected. At the same time, it may also need to be extended to specifically refer to aspects such as \textit{Email Address}, \textit{Email Contents}, or \textit{Attachments} which also can be used as \textit{instances} or be further specialised. This fits the SKOS modelling style better and also matches with how information is used by non semantic-web folks (e.g., to fill in forms), but does not fit well with semantic reasoning processes which largely use OWL (with Abox and Tbox assertions). DPV thus needs to satisfy both styles of using concepts as extensible instances and having support for OWL reasoning.

\subsection{Stakeholder Interoperability}

It is not possible for DPV to provide all relevant concepts in its on\-tol\-o\-gy/tax\-o\-no\-my given the almost infinite potential concepts (e.g., for personal data). Instead, DPV's community opted to provide the `most important' and `most commonly required' concepts, while requiring that DPV support stakeholders in extending DPV internally for reflecting the peculiarities of their own use-cases. For example, DPV can provide the purpose `Marketing' which a company can extend to describe `Summer Sale Offers'. 

Following from this, DPV's mission to provide interoperability across stakeholders relies on the `common' concept present in the DPV taxonomy as the basis for establishing shared understanding even if each stakeholder ends up creating their own unique or individual ontological representation. In the above example, \textit{Summer Sale Offers} may be incomprehensible to another stakeholder, but the use of DPV enables both entities to correctly interpret that this is a form of \textit{Marketing} — and thus be able to identify requirements and obligations for legal compliance associated with this concept. 

\section{Comparison with Related Work}\label{sec:related}
DPV fills an unique and necessary niche within the state of the art by providing concepts to represent legally relevant information related to the processing of personal data and use of technologies. Here we only describe related works that have a comparable adoption, or are standards, or are outputs of larger projects.

The Open Digital Rights Language (ODRL) \cite{ianella2007open} is a W3C standard for modelling policies and agreements which can dictate the permissions, prohibitions, and duties associated with use of data or resources. The ODRL vocabulary also provides specific concepts such as `obtain consent' to support commonly used agreements. While there is some overlap between DPV and ODRL, they are complimentary in their objectives and uses. ODRL focuses solely on providing a language for representation of policies and does not focus on specific legal requirements or jurisdictions. DPV's concepts can thus provide the necessary legal and jurisdictional relevant information within ODRL policies. This has been explored and demonstrated by existing work \cite{esteves_odrl_2021}. The DPVCG intends to establish close collaboration with the ODRL CG through its common members by aligning concepts between DPV and ODRL, and providing guidance for using DPV as an ODRL profile. DPV has also been mentioned by the International Data Spaces Association (IDSA) as a vocabulary of interest to support the implementation of GDPR and regulations in their policies based on ODRL \cite{bader2020international}.

LegalRuleML\footnote{\url{https://docs.oasis-open.org/legalruleml/legalruleml-core-spec/1.0.0/os/legalruleml-core-spec-1.0.0-os.html}} is an OASIS standard providing a `rule interchange language' that enables modelling and reasoning tasks based on legal arguments. Similar to ODRL, it focuses on providing a language for expression of `rules' and does not provide any taxonomies or model regulations. Similar to ODRL, DPV is also complimentary to LegalRuleML as a vocabulary that can be used to support representation of legal information. There is existing work that has explored the use of LegalRuleML for modelling requirements from the text of GDPR~\cite{robaldo2020dapreco}, though it has not been maintained nor been extended to support practical requirements related to implementing GDPR.

Gist\footnote{\url{https://www.semanticarts.com/gist/}} is a `minimalist upper ontology for the enterprise' which provides business concepts with a focus on minimising ambiguity. The Financial Industry Business Ontology\footnote{\url{https://spec.edmcouncil.org/fibo/}} (FIBO) models concepts relevant in financial business applications, such as contracts and financial transactions. Both Gist and FIBO represent `industrial ontologies' and have been developed over a significant portion of time with the involvement of corporate stakeholders. While neither support specific legal requirements or jurisdictions, Gist provides concepts relevant for modelling details about an organisation and FIBO provides modelling of contracts and contract-related processes — which DPV does not contain. Work is underway in DPVCG to study the FIBO contract concepts and identify how DPV can support contracts related to personal data processing.

\section{Adoption of DPV 1.0}\label{sec:adoption}

\subsection{Analysis of Citations}\label{sec:adoption-citations}

In this Section, the adoption of DPVCG's outputs is evaluated through a citation analysis performed over the DPV 1.0 publication~\cite{panetto_creating_2019}.
In this context, 81 publications that cite DPV were found through the Google Scholar service.
The gathered results underwent a review and were included in this analysis if deemed pertinent.
Duplicated publications, publications without an open-access version and in languages other than English were excluded from this analysis, resulting in 76 publications to be reviewed.
Table~\ref{tab:citation} presents the results of the performed publication evaluation.
The publications were evaluated in relation to their use of DPV: publications that reference DPV as a state of the art resource are signalled in the \textbf{Mention} column, that use DPV towards an application or use case in the \textbf{Use} column, and that extend DPV in the \textbf{Ext.} column.
Publications that contributed their extensions back to DPV (\textbf{Contrib.} column) are also marked in Table~\ref{tab:citation}, as well as if the work is applied to a certain domain or sector (\textbf{Domain} column). The \textbf{Effort} column denotes the amount of work (speculated) to update the implementation to DPV 2.0, where \texttt{++} denotes more work as implementations use a pre-1.0 version and will need to update the IRIs, \texttt{+} denotes minor efforts to check changed concepts in DPV 2.0, and \texttt{-} denotes no changes. DPV 2.0 is largely compatible with DPV 1.0, which means most implementations using DPV 1.0 can update with minimal changes.


\begin{table}[ht]
    \centering
    \caption{Citation analysis of academic publications that reference DPV 1.0~\cite{panetto_creating_2019}.}
    \label{tab:citation}
    \begin{tabular}{|l|c|c|c|c|c|c|c|}
        \hline
        \textbf{Work} & \textbf{Year} & \textbf{Mention} & \textbf{Use} & \textbf{Ext.} & \textbf{Contrib.} & \textbf{Domain} & \textbf{Effort} \\
        \hline
        \cite{calvaresi_agentbased_2020} & 2020 & X &  &  &  & Health & N/A \\
        \hline
        \cite{lieber_ecodalo_2020} & 2020 & X &  &  &  & Media & N/A \\
        \hline
        \cite{bonatti_machine_2020,matulevicius_method_2020,thalhath_metaprofiles_2020} & 2020 & X &  &  &  & & N/A \\
        \hline
        \cite{calvaresi_personal_2020,krasnashchok_towards_2020,leone_role_2020,ryan_design_2020} & 2020 &  & X &  &  & & ++ \\ 
        \hline
        \cite{calbimonte_decentralized_2020} & 2020 &  & X &  &  & Health & ++ \\
        \hline
        \cite{debruyne_justtime_2020,pandit_representing_2020,ryan_common_2020} & 2020 &  &  & X & X & & ++ \\
        \hline
        \cite{esteves_challenges_2021,mcdonald_evaluation_2021,flesch_investigating_2021} & 2021 & X &  &  &  & Health & N/A \\
        \hline
        \cite{grunewald_tira_2021,bonatti_representing_2021,sion_overview_2021,pandit_chapter_2021,hamed_enhancing_2021} & 2021 & X &  &  &  &  & N/A \\
        \hline
        \cite{ryan_gdpr_2021,esteves_odrl_2021,ekaputra_semanticenabled_2021,ryan_building_2021,leone_legal_2021} & 2021 &  & X &  &  & & ++ \\
        \hline
        \cite{garcia_towards_2021} & 2021 &  & X &  &  & Smart products & + \\
        \hline
        \cite{hickey_gdpr_2021} & 2021 &  &  & X & X &  & + \\
        \hline
        \cite{human_data_2022,jesus_consent_2022,human_advanced_2022,sangaroonsilp_mining_2022,pandit_proposals_2022,rasmusen_increasing_2022,esteves_semantifying_2022} & 2022 & X &  &  &  &  & N/A \\
        \hline
        \cite{esteves_analysis_2022} & 2022 & X &  &  & X &  & N/A \\
        \hline
        \cite{ryan_support_2022,ryan_dpcat_2022,pandit_semantic_2022,esteves_fostering_2022} & 2022 &  &  & X & X &  & + \\
        \hline
        \cite{becker_secondary_2022} & 2022 & X &  &  &  & Health & N/A \\
        \hline
        \cite{esteves_using_2022,debackere_policyoriented_2022,debackere_enforcing_2022,gambarelli_privaframe_2022,kurteva_making_2022} & 2022 &  & X &  &  &  & + \\
        \hline
        \cite{becher_contra_2022} & 2022 &  & X &  &  & IoT & + \\ 
        \hline
        \cite{hernandez_tikd_2022} & 2022 &  & X &  &  & Health & + \\
        \hline
        \cite{pandit_making_2023,bushati_what_2023,breteler_flint_2023,kurteva_relevant_2023,asgarinia_who_2023} & 2023 & X &  &  &  &  & N/A \\
        \hline
        \cite{gambarelli_is_2023,grunewald_enabling_2023,bailly_prototyping_2023,tang_helping_2023,esteves_towards_2023,esteves_using_2023,taheri_compliance_2023} & 2023 &  & X &  &  &  & + \\
        \hline
        \cite{sun_citizencentric_2023,navarrogallinad_evaluating_2023,florea_is_2023,gallinad_usable_2023} & 2023 &  & X &  &  & Health & + \\
        \hline
        \cite{kurteva_smashhitcore_2023} & 2023 &  & X &  &  & Smart cities & + \\
        \hline
        \cite{gupta_oppo_2023,zichichi_decentralized_2023} & 2023 &  & X &  &  & Media & + \\
        \hline
        \cite{esteves_semantics_2023,pandit_towards_2023} & 2023 &  &  & X & X &  & + \\
        \hline
        \cite{raza_semantic_2023} & 2023 &  &  & X &  & Health & + \\
        \hline
        \cite{pandit_enhancing_2024} & 2024 &  & X &  &  & Health & - \\
        \hline
        \cite{herwanto_leveraging_2024} & 2024 &  & X &  &  &  & + \\
        \hline
        \cite{golpayeganiAIROOntologyRepresenting2022,golpayeganiBeHighRiskNot2023,golpayeganiAICardsApplied2024} & 2024 & & X & X & WIP & EU AI Act & ++ \\
        \hline
        \cite{panditImplementingISOIEC2024} & 2024 & & X & & & & - \\ 
        \hline
    \end{tabular}
\end{table}

In this context, DPV's specifications were compared against other state of the art vocabularies in the data protection domain regarding their ability to represent information related to GDPR rights and duties~\cite{esteves_analysis_2022}, their machine-readability, maintenance, accessibility, GDPR support and existence of compliance tools~\cite{asgarinia_who_2023}, and their capacity to aid with data interoperability and adhere to the FAIR principles~\cite{thalhath_metaprofiles_2020}.
In all mentioned surveys, DPV obtained a higher score compared with other existing solutions.
When it comes to extensions performed over DPV, most were contributed back to DPV to be integrated into DPVCG's outputs.
Concerning work on GDPR requirements, there were proposed extensions focusing on consent~\cite{pandit_representing_2020,debruyne_justtime_2020}, in particular related to the processing of electronic health record data~\cite{raza_semantic_2023}, as well as on building semantic models to represent records of processing activities~\cite{ryan_common_2020,ryan_support_2022,ryan_dpcat_2022,ryan_building_2021}, data protection impact assessments~\cite{pandit_semantic_2022}, data breaches' reports~\cite{pandit_towards_2023}, and international data transfer notices~\cite{hickey_gdpr_2021}.
Moreover, extensions focusing on GDPR's data subject rights and exemptions to these rights~\cite{esteves_fostering_2022} and on DGA requirements~\cite{esteves_semantics_2023,esteves_towards_2023} were also contributed back to DPVCG's outputs.

In terms of applications to specific use cases, there is a body of work focusing on providing tools for auditing and GDPR compliance evaluation~\cite{ryan_design_2020,ryan_gdpr_2021,taheri_compliance_2023}, on data minimisation~\cite{garcia_towards_2021}, as well as on the documentation and annotation of privacy policies~\cite{leone_role_2020,leone_legal_2021,krasnashchok_towards_2020,gupta_oppo_2023} and privacy preferences~\cite{becher_contra_2022}.
Moreover, ML models were trained with DPV's taxonomies to identify personal data processing activities in code repositories~\cite{tang_helping_2023,herwanto_leveraging_2024} and textual datasets~\cite{gambarelli_privaframe_2022,gambarelli_is_2023}.
DPV's outputs were also used to model access and usage control policies~\cite{calvaresi_personal_2020,calbimonte_decentralized_2020,ekaputra_semanticenabled_2021,zichichi_decentralized_2023}, and in particular applied to Solid~\cite{esteves_odrl_2021,esteves_using_2022,debackere_policyoriented_2022,debackere_enforcing_2022,florea_is_2023,bailly_prototyping_2023,esteves_using_2023} and health data-sharing use cases~\cite{sun_citizencentric_2023,pandit_enhancing_2024}, as well as to describe consent records and contracts for sensor data~\cite{kurteva_smashhitcore_2023,kurteva_making_2022}.
In the context of data spaces, DPV was used to provide descriptions of health data handling activities~\cite{hernandez_tikd_2022} and to create user-centric privacy interfaces~\cite{grunewald_enabling_2023,navarrogallinad_evaluating_2023,gallinad_usable_2023}.

\subsection{Projects and Industrial use of DPV}\label{sec:adoption-project}

DPV and DPVCG are outputs of the SPECIAL H2020 project and were actively developed and used within the project. In addition to this, DPV was also used in TRAPEZE, MOSAICrOWN, smashHit, FAIRVASC, and PROTECT ITN projects funded under the EU's H2020 programme which involved both academic and commercial partners. In addition, SPECIAL and TRAPEZE also included a Data Protection Authority who provided legal expertise in implementation and design of ontologies. TRAPEZE actively contributed back to the DPV and was instrumental in identifying the design structure where both RDFS+SKOS and OWL serialisations were developed, and supported development of a multilingual documentation framework to be implemented in future versions. 

In addition to the use of DPV in industrial context in the above projects,~companies that actively utilise DPV include Signatu\footnote{\url{https://signatu.com/}} — which develops legal compliance solutions, JLINC\footnote{\url{https://www.jlinc.com/}} which develops digital data agreements, and Inrupt\footnote{\url{https://www.inrupt.com/blog/the-benefits-of-dynamic-user-consent}} which develops Solid specifications and implementations. The DPVCG has received contributions from these companies, with Signatu being an active contributor by providing legal expertise and requirements for industrial applications.

\subsection{Use of DPV in Standards}\label{sec:adoption-standards}

ISO/IEC TS 27560:2023 is a Technical Specification (TS) describing a `consent record information structure', which defines the specific information to be maintained in consent records. DPV's consent modelling played a significant part in the development of this standard based on sharing the knowledge
of legal and semantic requirements regarding consent records and receipts. The annex of 27560:2023 provides examples of consent records and receipts using JSON-LD where DPV is explicitly referenced in the document and its concepts are used as to define the schema (e.g., \textit{hasPurpose} and instances (e.g., \textit{Service Provision}).

The IEEE P7012 Working Group is developing a specification to define how ``personal privacy terms are proffered and how they can be read and agreed to by machines''. It explicitly references DPV as the vocabulary to describe activities regarding processing of personal data in an interoperable manner.

\section{Data Privacy Vocabulary 2.0}\label{sec:overview}

\begin{figure*}[ht]
\caption{Overview of concepts in DPV and their extensions}
\label{fig:dpv}
\centering
\fbox{\includegraphics[width=\textwidth]{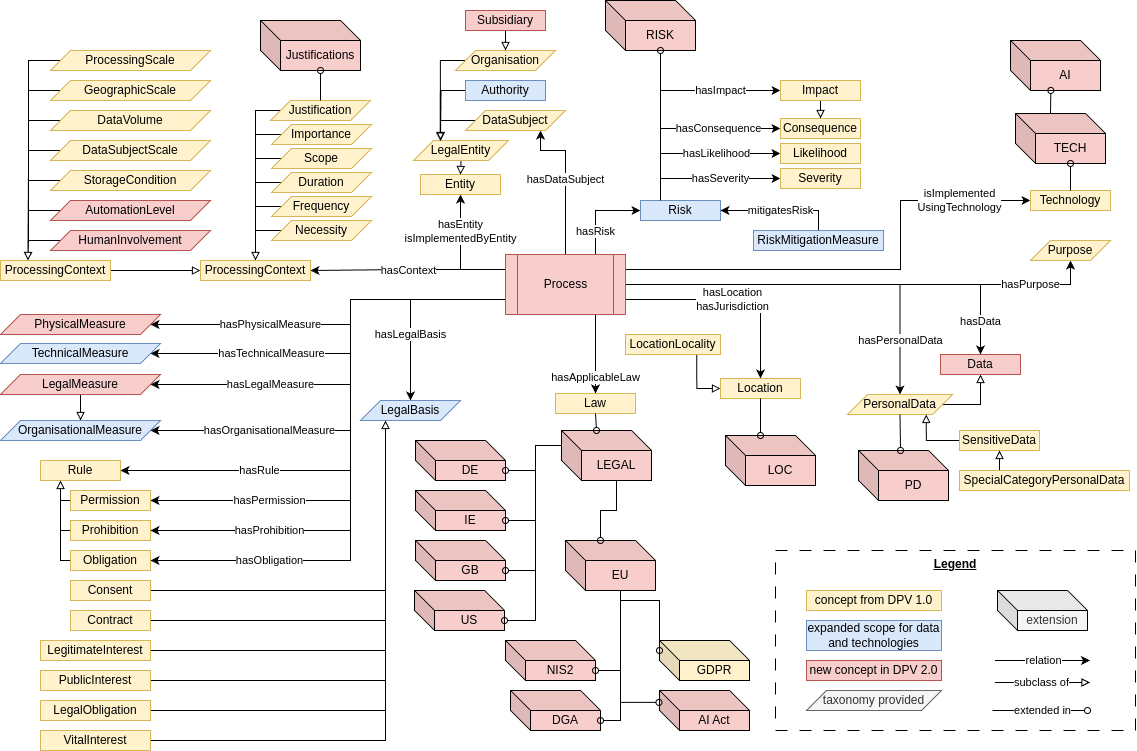}}
\end{figure*}

The motivation of the Data Privacy Vocabulary (DPV) is to provide a `data model' or a `taxonomy' of concepts that act as a vocabulary for the interoperable representation and exchange of information about personal data and its processing. For this, the DPV specification represents an abstract model of concepts and relationships that can be implemented and applied using technologies appropriate to the use-case's requirements. 
The core concepts in DPV, as shown in Figure \ref{fig:dpv}, are broadly as follows:
\begin{itemize}
    \item \textbf{Purpose}: end-goal for why personal data is processed, e.g., Service Provision
    \item \textbf{Processing}: representing operations over personal data, e.g., Collect, Store
    \item \textbf{Personal Data}: categories of personal data involved
    \item \textbf{Legal Basis}: justification in law for performing this activity
    \item \textbf{Legal Roles}: Data Controller, Data Subject, etc.
    \item \textbf{Technical and Organisational Measures}: safeguarding activities
    \item \textbf{Processing Context}: storage conditions, automation, scale and scope
    \item \textbf{Context}: concepts other than the above, e.g., necessity, duration
    \item \textbf{Rights}: legally recognised rights associated with activities
    \item \textbf{Risk and Risk Mitigation}: managing risks, consequences, and impacts
\end{itemize}

Each of these `core concepts' are expanded into taxonomies to reflect their application in use-cases. The taxonomies also provide `knowledge' by asserting categorisation based on the core concepts. For example, \textit{Personal Data} is a core concept which is specialised into \textit{Sensitive Personal Data}, and the taxonomy expanding upon personal data contains some instances asserted as being sensitive. An adopter therefore not only gets a taxonomy of personal data, but is also able to utilise the categorisation of it into sensitive personal data.

To assist newcomers in understanding the structure of DPV and how its concepts are organised — a \textit{Primer}\footnote{\url{https://w3id.org/dpv/primer}} document has been developed. In addition, the documentation is continually refined to provide illustrative guidance and examples\footnote{\url{https://w3id.org/dpv/2.0/examples}}, and a searchable index\footnote{\url{https://w3id.org/dpv/2.0/search}} of concepts is also provided. The below example shows the use of DPV in implementing consent records as used in ISO/IEC TS 27560:2023 Privacy technologies — Consent record information structure~\cite{panditImplementingISOIEC2024}, where this work was recently also presented to the EU Commission:
\begin{minted}[fontsize=\footnotesize]{turtle}
@prefix dct: <http://purl.org/dc/terms/> .
@prefix dpv: <https://w3id.org/dpv#> .
@prefix loc: <https://w3id.org/dpv/loc#> .
@prefix eu-gdpr: <https://w3id.org/dpv/legal/eu/gdpr#> .
@prefix : <https://example.com/> .
:63ded36f-4acd-4f3c-991e-6cb636698523 a dpv:ConsentRecord ;
    dct:hasVersion "ISO-27560" ;
    dpv:hasIdentifier "63ded36f-4acd-4f3c-991e-6cb636698523" ;
    dpv:hasDataSubject "96121fde-199f-4848-8942-4436e270513a" ;
    dpv:hasNotice <https://example.com/notice> ;
    dpv:hasProcess [
        a dpv:Process ;
        dct:title "Send Newsletters with Seasonal Offers"@en ;
        dpv:hasPurpose dpv:Marketing ;
        dpv:hasLegalBasis dpv:Consent, eu-gdpr:A6-1-a ;
        dpv:hasPersonalData pd:Email ;
        dpv:hasDataController ex:Acme ;
        dpv:hasProcessing dpv:Collect, dpv:Store ;
        dpv:hasStorageCondition [ 
            dpv:hasLocation loc:IE ;
            dpv:hasDuration "P1Y"^^xsd:duration ; ] ;
        dpv:hasJurisdiction loc:EU ;
        dpv:hasRecipient :Beta, :Epsilon ; ] ;
    dpv:hasConsentStatus dpv:ConsentGiven ;
    dct:hasPart [
        a dpv:ConsentGiven, dpv:ExplicitlyExpressedConsent ;
        dpv:isIndicatedAtTime "2021-05-28T12:24:00"^^xsd:dateTime ;
        dpv:hasDuration "P1Y"^^xsd:duration ;
        dpv:hasEntity "96121fde-199f-4848-8942-4436e270513a" ] .
\end{minted}

Extensions are a collection of concepts provided in a separate namespace. They are used to represent specific concepts in jurisdictions and regulations, for example \textit{Consent} is present in the `main' or `core' DPV, and is expanded upon in the \textit{EU-GDPR} extension to represent the specific requirements for consent under GDPR. Extensions are also used to provide a large group of concepts for a specific topic as their inclusion in the main vocabulary would not be practical or would introduce ambiguities between concepts. For example, the \textit{Personal Data} extension provides a taxonomy of personal data categories, which were taken out of the main vocabulary due to ambiguity and confusion in concepts such as \textit{Location} being used for both personal data and data storage location.

A list of ongoing work in DPVCG is as follows:
\begin{itemize}
    \item Data Privacy Specification (DPV) -- \url{https://w3id.org/dpv}
    \item Personal Data Concepts extension (PD) -- \url{https://w3id.org/dpv/pd}
    \item Location concepts extension (LOC) -- \url{https://w3id.org/dpv/loc}
    \item Legal concepts extension (LEGAL) -- \url{https://w3id.org/dpv/legal}
    \item EU GDPR concepts (EU-GDPR) -- \url{https://w3id.org/dpv/legal/eu/gdpr}
    \item EU DGA concepts (EU-DGA) -- \url{https://w3id.org/dpv/legal/eu/dga} 
    \item EU AI Act concepts (EU-AIAct) -- \url{https://w3id.org/dpv/legal/eu/aiact}
    \item Legal Concepts for Ireland, Germany, United Kingdom, USA -- \url{https://w3id.org/dpv/legal/IE} (Replace \textit{IE} with ISO 3166-1 code, e.g., IE/DE/GB/US)
    \item Risk Management Concepts (RISK) -- \url{https://w3id.org/dpv/risk}
    \item Technology Concepts (TECH) -- \url{https://w3id.org/dpv/tech}
    \item AI Technology Concepts (AI) -- \url{https://w3id.org/dpv/ai}
    \item Justifications -- \url{https://w3id.org/dpv/justifications}
\end{itemize}

\subsection{Major Changes in 2.0}
DPV 2.0 and all its extensions contain 2394 concepts (with 2198 classes and 196 properties), with 1017 concepts added and 805 concepts removed as compare to DPV 1.0. In DPV 1.0, only DPV, Personal Data (PD), and the EU-GDPR extension were provided as `complete', with the others specified to be in \textit{draft} mode. In DPV 2.0, DPV along with all of its extensions of PD, LOC (locations), LEGAL (including jurisdictional laws such as EU GDPR), RISK, TECH, AI, and Justifications have been provided as finalised resources. A detailed changelog\footnote{\url{https://w3id.org/dpv/2.0/changelog}} has been provided expanding on information in this section, including added/removed concepts in DPV and extensions.

\noindent\textbf{Change in scope:} The scope of concepts in DPV 1.0 was limited to `processing of personal data'. In DPV 2.0, the scope was expanded to include `any data or technology' to have the same semantic structure for management of both personal and non-personal data and technologies (including AI). This enables DPV to support regulations such as the Data Governance Act (DGA) which motivates `reuse of personal and non-personal data' and the AI Act where existing DPV concepts such as \textit{Purpose}, \textit{Rights}, and \textit{Risk} can be reused. While the scope of the DPVCG is still limited to personal data (and associated technologies), the expansion of scope for concepts enables the DPV to be utilised for a much broader range of use-cases and regulations. More importantly, it provides a common mechanism for representing information about activities in the so called `AI and Data' regulations, and makes alignments with existing standards such as ODRL easier to manage. The expansion of scope is backwards compatible with DPV 1.0 as it does not change the application and interpretation of concepts.

\noindent\textbf{Change in semantics:} DPV 1.0 was provided with three different semantics: a custom extension of SKOS as the `default' along with RDFS+SKOS and OWL2 variants — each with a distinct namespace. In DPV 2.0, the custom SKOS extension has been removed and replaced with RDFS+SKOS as the default with an OWL2 variant — each with a distinct namespace. This change is largely backwards-compatible as both DPV 1.0 and 2.0 use \texttt{skos:Concept} with the IRI for DPV 1.0 default and RDFS+SKOS redirected to DPV 2.0 RDFS+SKOS namespace, and that for DPV 1.0 OWL2 to DPV 2.0 OWL2 namespace.

\noindent\textbf{Versioned IRIs:} DPV 1.0 utilised unversioned IRIs (e.g., \texttt{w3id.org/dpv} — which is not considered best practice. DPV 2.0 introduces versioned IRIs to enable distinguishing between versions and choosing a specific version to use regardless of future changes. The versioned IRI for DPV 1.0 is \texttt{w3id.org/dpv/1.0} and that for DPV 2.0 is \texttt{w3id.org/dpv/2.0}. Extension namespaces are constructed by suffixing the versioned DPV namespace, e.g., \texttt{w3id.org/dpv/2.0/pd}. The unversioned IRIs redirect to the latest DPV version.

\noindent\textbf{Change in extensions:} 
In addition to introduction of new extensions, DPV 2.0 also changes the namespaces and management of extensions. In DPV 1.0, the \texttt{dpv-pd}, \texttt{dpv-legal}, \texttt{dpv-gdpr}, \texttt{dpv-nace}, and \texttt{dpv-tech} extensions used the prefix \texttt{dpv-} in their folder structure and namespaces whereas \texttt{risk} and \texttt{rights} did not. 
In DPV 2.0, extensions are defined without the prefix for consistency. 

The \texttt{dpv-legal} extension in DPV 1.0 was a \textit{draft} providing legal concepts (laws, authorities) and locations based on ISO 3166-1 codes.
In DPV 2.0 it has been split into \texttt{legal} and \texttt{loc} (locations) extensions for separation of concerns. Both legal and location extensions are provided as completed in DPV 2.0. Further, the location concepts have been aligned with EU Vocabularies\footnote{\url{https://op.europa.eu/s/zQhL}} as an example of connecting DPV locations to external vocabularies. 

The namespaces and organisation of legal concepts in DPV 2.0 has been redesigned to distinguish between jurisdictions and laws by using the ISO 3166-1 codes to create a structured path. For example, the namespace for EU-GDPR extension is \texttt{w3id.org/dpv/legal/eu/gdpr} — which reflects that it is a legal extension associated with EU jurisdiction and models the GDPR regulation. This mechanism also enables laws with the same name in other jurisdictions to be declared without conflicts e.g. UK's GDPR would be under the namespace \texttt{/legal/gb/gdpr}. And it keeps all laws associated with a specific jursidiction within the same path, e.g., DGA, NIS2, and AI Act are represented within the \texttt{/legal/eu} namespace as EU laws. The \textit{draft} EU Rights extension in DPV 1.0 providing concepts from EU Charter of Fundamental Rights has been moved to \texttt{/legal/eu/rights} namespace in DPV 2.0 following this reorganisation.

The extension \texttt{dpv-nace} modelled the NACE\footnote{\url{https://ec.europa.eu/eurostat/web/nace}} 2.0 taxonomy of economic activities provided by the EU as RDFS concepts for use with OWL vocabularies as the EU uses SKOS to declare NACE concepts. The extension has been removed in DPV 2.0 as NACE has recently been updated to version 2.1, and the extension did not provide any meaningful benefit and increased maintenance cost. To the best of our knowledge, the extension was not being used, and the recommendation is to use the authoritative NACE taxonomy going ahead.

The expected impact of these changes for DPV-GDPR and DPV-PD in DPV 1.0 is minimal as their unversioned IRIs are redirected to DPV 2.0 which contains the same concepts. For draft extensions in DPV 1.0, there are breaking changes — most severely in the case of legal/location concepts due to their separation into two extensions. In case an adopter has been using these draft extensions without being aware of impending changes, we estimate a minimal effort to use the new (and improved) DPV 2.0 extensions instead. In any case, with the versioned IRIs the adopters can continue use of DPV 1.0 if desired.

\noindent\textbf{Changes in DPV Concepts:} Of the 911 concepts in DPV 2.0, 311 are new additions and 56 concepts were removed as compared to DPV 1.0. The removed concepts represent refinements
and moving concepts to an extension, e.g., harms and other impacts were moved to the RISK extension.

\section{Methodology \& Design Principles}\label{sec:design}

\subsection{Management by DPVCG} 

The DPVCG used the W3C infrastructure to manage development of DPV, which consisted of the mailing list, task management, and namespace management (\textit{w3.org}). After the migration of W3C to GitHub, DPVCG utilised the provided repository\footnote{\url{https://github.com/w3c/dpv/}} for version control, task management, discussions, and contributions.
In its meetings, the group utilised spreadsheets (using Google Sheets for collaboration) to support the (lack of) technical knowledge of members and ease of discussion, commenting, and sharing. Formal discussions and approvals were undertaken via the mailing list and meetings.

\subsection{Ontology Engineering Process}

The DPVCG consists of experts from multiple disciplines — computer science, law, sociology, authorities, and others. The primary role of its members is to discuss and reach consensus on the scope and information to be represented in DPV. Ontology engineers are then responsible for providing the appropriate modelling of concepts and organisation of DPV as a semantic web resource. While the DPVCG did not formally establish an ontology engineering methodology, practices that were adopted and evolved in the community reflect commonly utilised engineering methodologies of NeOn \cite{suarez2011neon} and LOT \cite{poveda2022lot}.

The development process generally contained a member of the group proposing addition or modification of a concept, with information shared potentially via the mailing list and/or GitHub repo,
and discussions and decisions within the meetings. Domain experts offered their advice on the information being modelled and what aspects of this should be considered, which were then formalised and shared as proposals. The group then discussed and voted to resolve the proposal, with minutes of the meeting reflecting the discussions and resolutions. 

Compared to formal ontology engineering methods, this ad-hoc approach lacked the explicit documentation of requirements, competency questions, and use-cases — which would have been expensive to maintain given the limited (regular) participation of volunteers. However, given the stability of the group and continued iteration and adoption of DPV, the group has identified this as an important step to undertake in the future.

The data in spreadsheets was structured to support ontology/taxonomy creation while still being comprehensible to the non-semantic web members. For generating the RDF serialisations, a custom documentation generator was developed which downloaded the spreadsheets
and serialised them into multiple RDF formats, and generated the corresponding HTMLs. Existing tools such as WIDOCO \cite{garijo2017widoco} were not used due to their limitation in control of outputs — for example DPV has multiple taxonomies
which would be all grouped together in a giant list in WIDOCO outputs, and which in DPV outputs are grouped separately to support adopters seeing related concepts grouped together. WIDOCO also did not support the dual RDFS+SKOS and OWL modelling outputs of DPV, or the ReSpec template\footnote{\url{https://respec.org/docs/}} common in W3C outputs. 

DPV follows best practices and guidelines established within the semantic web community. Namely, the W3C Best Practices for Publishing Linked Data (2014)
WIDOCO best practices \cite{garijo2017widoco} with OOPS!\footnote{\url{https://oops.linkeddata.es/}} and FOOPS!\footnote{\url{https://foops.linkeddata.es/}} for evaluation, W3ID\footnote{\url{https://w3id.org/}} for permanent IRIs, GitHub for version control and collaboration, and Zenodo for archival.

\subsection{Implications of Using SKOS and OWL}

As described in the requirements in Section \ref{sec:req}, use-cases utilising DPV involve cases where its concepts are used as instances (taxonomy) or as a schema that is instantiated (ontology). Initially, DPV was only provided as an OWL2 ontology. This was expanded upon in DPV 1.0 which used custom SKOS extensions to define the `base' vocabulary with serialisations in RDFS+SKOS and OWL2 with the goal of supporting both categories of use-cases. In DPV 2.0, the custom SKOS extension was removed in favour of using RDFS+SKOS as standards for the default serialisation and providing an alternative serialisation for OWL2. 

\begin{figure*}[ht]
\caption{Semantic implications of using DPV with RDFS+SKOS and OWL semantics}
\label{fig:semantic-implications}
\centering
\fbox{\includegraphics[width=\textwidth]{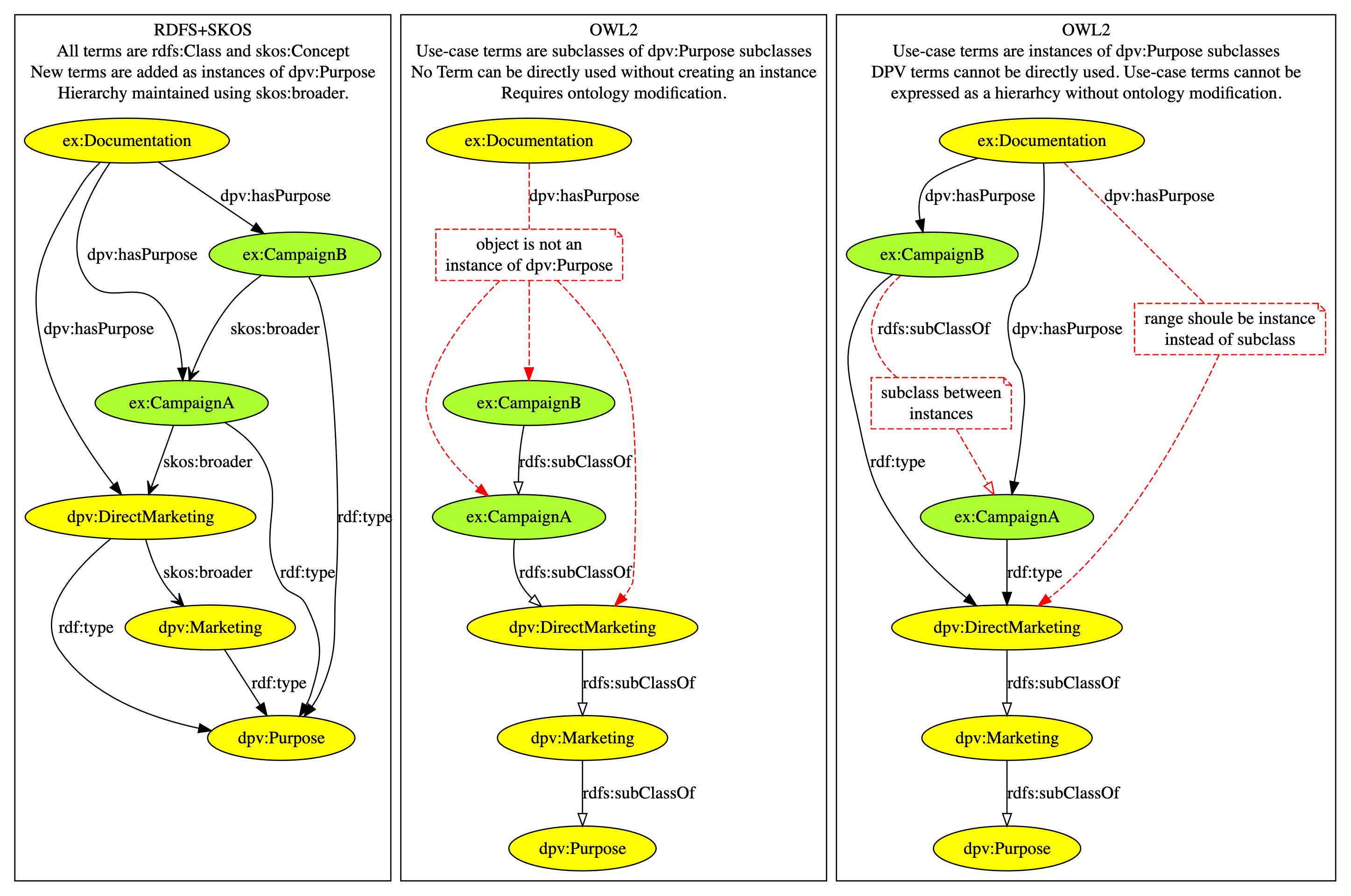}}
\end{figure*}

The RDFS+SKOS serialisation defines concepts as instances of \texttt{rdfs:Class} and \texttt{skos:Concept}. To create a hierarchical taxonomy, the concepts are represented as instances of a top-concept (e.g., \texttt{dpv:Marketing} as an instance of \texttt{dpv:Purpose}) and \texttt{skos:broader/narrower} is used to define the relation between instances. In OWL2, concepts are defined as \texttt{owl:Class} and the hierarchy is defined using \texttt{rdfs:subClassOf}. The justification for why DPV is provided with two different semantics is illustrated in Figure~\ref{fig:semantic-implications}.

The Figure compares the implementation of RDFS+SKOS and OWL2 for an use-case that uses the purpose taxonomy from DPV. The use-case involves three steps of documentation (not shown in the figure) where the organisation first records that its planned purpose is `Direct Marketing', and then later it creates `CampaignA' as a specific form of direct marketing, and even later creates `CampaignB' as a specific part of `CampaignA' direct marketing.

In RDFS+SKOS, both \texttt{dpv:DirectMarketing} and \texttt{ex:CampaignA} are both defined as an instance of \texttt{dpv:Purpose}, and associated with each other  using \texttt{skos:broader}. Using these as the object with property \texttt{dpv:hasPurpose} is correct as its range is (instances of) \texttt{dpv:Purpose}. Later, when \texttt{ex:CampaignB} is introduced, it does not require changes to DPV or the use-case graph (e.g., to convert instances into classes) as the new concept is also defined as an instance of \texttt{dpv:Purpose} and can be used with the property.

In OWL2, all DPV concepts are defined as instances of \texttt{owl:Class} and associated with each other using \texttt{rdfs:subClassOf}. The use-case concepts can now be declared as either subclasses of DPV concepts or as instances. In either case, DPV concepts cannot be directly used with the property \texttt{dpv:hasPurpose} as they are not instances of \texttt{dpv:Purpose} (shown in red). If the use-case creates instances — as shown where \texttt{ex:CampaignA} is created as an instance of \texttt{dpv:DirectMarketing} — then the use-case concept can be correctly used with \texttt{dpv:hasPurpose}. However, when \texttt{ex:CampaignB} is later introduced, it cannot be subclass of \texttt{ex:CampaignA} as `subclass of instance' is undefined\footnote{It might be tempting to use punning here — but it would not be correct
as we have a concept as class and instance in the same context — which is undefined.} in OWL2. The relation between the two use-case concepts thus must be defined using either SKOS (thereby mixing SKOS and OWL2) or using another property, e.g., \texttt{dct:hasPart}. At this point, the use-case should reengineer its ontology by creating a class representing \texttt{ex:CampaignA} and creating the necessary subclass and instances to represent the relationship with \texttt{ex:CampaignB}. This however does not solve the issue with directly using DPV concepts as instances.

Thus, the RDFS+SKOS model is suitable for when DPV is to be used as a controlled taxonomy with a `lightweight ontology' that supports extending the taxonomy in use-cases. Conversely, the OWL2 model is better suited for cases where formal reasoning is needed, and where sufficient ontology engineering capabilities exist to address changes in use-cases. By providing both serialisations, DPV enables the adopters to choose the most suitable serialisation that supports their use-case and/or existing implementations, and retains semantic interoperability based on converting between SKOS and OWL2\footnote{Using OWL and SKOS - Sean Bechhofer and
Alistair Miles, W3C (2008)}



\section{Conclusion \& Future Work}\label{sec:conclusion}

The Data Privacy Vocabulary (DPV), in its second iteration, thus provides a significantly richer, extended, and state of the art resource which fills an important niche in the current landscape regarding expression of information associated with personal data processing and the use of technologies. This article highlighted the motivation for its development, described its methodology and design processes, and showcased its value evidenced through adoption across academia, industry, and standards. 

DPV version 2 represents a significant milestone in the development of a vocabulary to support legally relevant processes across multiple regulations and jurisdictions,  as well as recent advances in AI and data sharing regulations such as through EU's DGA and AI Act, and in architectures such as Solid and IDSA. The DPVCG, in continuing to develop the DPV, welcomes more participation and contributions to support its vision of providing an interoperable vocabulary that provides value and supports making legal compliance processes more efficient and aligned for all stakeholders.




The DPVCG plans to refine its TECH and AI extensions based on existing works \cite{golpayeganiAIROOntologyRepresenting2022,golpayeganiAICardsApplied2024} providing taxonomies for AI techniques, capabilities, lifecycle stages, risks and risk sources, and to enable stakeholders to express specific use-cases (e.g., involving generative AI) in a manner that supports requirements for EU AI Act and ISO standards \cite{golpayeganiAICardsApplied2024}. 
The DPVCG is also continuing its efforts to develop vocabularies to represent key `data and AI regulations' notably in EU the Digital Services Act (DSA), Data Markets Act (DMA), Data Act, and Data Spaces, and modelling laws in other jurisdictions, e.g., Ireland, USA, and UK.

To support the application of DPV in regulatory environments, the DPVCG is developing guides based on existing work that utilises DPV to support GDPR implementations. These include Records of Processing Activities (ROPA) \cite{ryan_dpcat_2022}, Data Protection Impact Assessment (DPIA) \cite{pandit_semantic_2022}, Data Breaches \cite{pandit_towards_2023}, and Rights management. To support the implementation of decentralised and user-centric applications of DPV, such as those envisioned in the IDSA, IEEE P7012 and Solid, the DPVCG is developing vocabularies and guides —  for example implementation of ISO Standards for (semantic) privacy notices. In addition, the DPV is also looking to incorporate the Standard Data Protection Model\footnote{\url{https://www.bfdi.bund.de/EN/Fachthemen/Inhalte/Technik/SDM.html}} provided by the German data protection authority regarding implementation of concrete technical and organisational measures.


\paragraph*{Resource Availability Statement:} The source and releases for the DPV are available via GitHub: \url{https://github.com/w3c/dpv} and have been deposited in Zenodo for long-term archival: \url{https://doi.org/10.5281/zenodo.12505840}.

\textbf{Acknowledgements of Funding}
\footnotesize
The DPVCG was established as part of the SPECIAL H2020 Project funded by the European Union’s Horizon 2020 programme under Grant\#731601 (2017-2019).
Harshvardhan J. Pandit was funded (2020-2022) by the Irish Research Council's Government of Ireland Postdoctoral Fellowship Grant\#GOIPD/2020/790. The ADAPT SFI Centre for Digital Media Technology is funded by Science Foundation Ireland through the SFI Research Centres Programme and is co-funded under the European Regional Development Fund (ERDF) through Grant\#13/RC/2106 (2018-2020) and Grant\#13/RC/2106\_P2 (2021 onwards). Piero Bonatti and Luigi Sauro were funded by the European Union’s Horizon 2020 programme under Grant\#731601 (2017-2019), and under TRAPEZE Grant\#883464 (2020-2023).
Beatriz Esteves and Delaram Golpayegani were funded by European Union’s Horizon 2020 programme's Marie Skłodowska-Curie Grant\#813497 PROTECT ITN Project.
Beatriz is also funded by SolidLab Vlaanderen (Flemish Government, EWI and RRF project VV023/10).
Julian Flake received funding from the German Federal Ministry of Education and Research (BMBF) grant\#16KIS1298 (AI-NET PROTECT), from the European Union's Horizon Europe Framework Programme grant\#101129822 (TITAN) and from the European Union's Digital Europe Programme grant\#101123471 (EDGE-Skills).
For the purpose of Open Access the authors have applied a CC-BY public copyright licence to any Author Accepted Manuscript arising from this submission.
\normalsize

\bibliography{paper}
\bibliographystyle{splncs04}

\end{document}